\documentstyle[preprint,tighten,aps,epsf,floats]{revtex}

\newcommand{\be}{\begin{equation}}
\newcommand{\ee}{\end{equation}}
\newcommand{\ba}{\begin{eqnarray}}
\newcommand{\ea}{\end{eqnarray}}

\def\vec#1{{\mbox{\boldmath$#1$}}}
\def\ket#1{\vert #1 \rangle}
\def\bra#1{\langle #1 \vert}

\newcommand{\e}{\mbox{$\vec{e}$}}

\begin{document}

\draft
\preprint{SLAC-PUB-8439,hep-lat/0004016}
\newcommand{\mar}{\marginpar{***}}

\title{The Lattice  Schwinger Model:  Confinement, Anomalies, Chiral
Fermions
and All That
\thanks{
Work supported by 
	Department of Energy contract DE-AC03-76SF00515.
}
}

\author{Kirill Melnikov\thanks{
e-mail:  melnikov@slac.stanford.edu}
and Marvin Weinstein \thanks{
e-mail:  niv@slac.stanford.edu}
}

\address{Stanford Linear Accelerator Center\\
Stanford University, Stanford, CA 94309}

\maketitle

\begin{abstract}
In order to better understand what to expect from numerical CORE
computations for two-dimensional massless QED (the Schwinger model)
we wish to obtain some analytic control over the approach to
the continuum limit for various choices of fermion derivative.  To this
end we study the Hamiltonian formulation of the lattice Schwinger model
(i.e., the theory defined on 
the spatial lattice with continuous time) in $A_0=0$ gauge.
We begin with a discussion of the solution of the Hamilton equations
of motion in the continuum, we  then parallel
the derivation of the continuum solution within the  lattice framework
for a range of fermion derivatives. The equations of motion for the
Fourier
transform of the lattice charge density operator show explicitly why
it is a regulated version of this operator which corresponds to the
point-split operator
of the continuum theory and the sense in which the regulated lattice
operator
can be treated as a Bose field.   The same formulas explicitly exhibit
operators
whose matrix elements measure the lack of approach to the continuum
physics.
We show that both chirality violating
Wilson-type and chirality preserving SLAC-type derivatives correctly
reproduce
the continuum theory and show that there is a clear
connection between the strong and weak coupling limits of a theory
based upon a generalized SLAC-type derivative.
\end{abstract}

\newpage

\section{Introduction}

It was argued in an earlier paper\cite{CORE} that the Contractor
Renormalization
Group(CORE) method can be used to map a theory of lattice fermions and
gauge fields
into an equivalent highly frustrated anti-ferromagnet.
Although explicit computations were presented only for the free fermion
theory,
it was argued that a corresponding mapping must exist for the interacting
theory because the space of {\it retained states\/} used for
the free theory coincides with the set of lowest energy states of the
strongly coupled gauge theory.  While this argument is true, it is
obviously
important to have a better understanding of the details of how the mapping
works.
In order to get some experience with this process
for a theory which is well understood we decided to study the lattice
Schwinger model (i.e., two-dimensional QED), since the exact continuum
solution
of this model exists.  Before diving into the CORE computation, however,
we first
needed to understand the degree to which the lattice model exhibits the
interesting
features of the continuum theory.  This paper is devoted to an analytical
treatment
of the lattice Schwinger model with an eye to clarifying the physics which
underlines
the continuum solution and identifying those general features of the model
which
should provide an ultimate check of the correctness of {\it any\/}
numerical solution.

The continuum Schwinger
model\cite{Schwinger,Lowenstein,KogutandSusskind,Coleman,Manton,Murayama},
in addition to being a non-trivial interacting theory of fermions and
gauge fields, provides a laboratory for studying a wide range of
interesting
phenomena.  It exhibits: confinement of the fermionic degrees
of freedom and the concomitant appearance of a massive boson
in the exact spectrum; breaking of chiral symmetry through the
axial anomaly; screening of external charges and background electric
fields; infinite degeneracy of the vacuum states of the theory
(theta parameters);
and the ability to produce arbitrary fermionic polarization charge
densities
by applying an operator of the form $e^{i \int {\rm d}x\, \alpha(x)
j_5^0(x) }$
to the vacuum state (due to the anomalous commutator of the electric and
axial-charge density operators).  It is important to ask which
of these features can be understood in the lattice theory before taking
the continuum limit and how complicated a CORE computation has to be in
order to extract this physics. Although the literature contains
discussions
of various aspects of the model, such as confinement and the
axial anomaly\cite{Susskind,Ambjorn,Bodwin,Nason:1985yp},
we are not aware of any
systematic discussion of the theory which attempts to parallel
the derivation of the continuum solution within the lattice framework.
This is what we do in this paper.

In order to make the physics as transparent as possible we formulate
the Hamiltonian version of the theory in $A_0 = 0$ gauge and only then
rewrite it
within the super-selected sector of {\it gauge-invariant states\/}.
We then study the Hamilton equations of motion for the electric charge
density
operator, whose form is completely determined by the way in which local
gauge
invariance is introduced into the lattice theory.  Obviously, the form
of the operator equations of motion depends upon the specific
lattice fermion derivative and so we study this problem for a wide class
of different derivatives; in particular, generalizations of the so-called
SLAC
derivative\cite{slacderivative}, which explicitly maintain the lattice
chiral
symmetry and generalizations of the Wilson derivative\cite{Wilson},
which break the chiral symmetry for non-zero momenta.  We find that all
of these approaches produce a satisfactory treatment of the continuum
theory,
however the detailed physical picture of how things work varies
greatly.

We show that a key
issue for connecting the lattice theory to the continuum theory is
which lattice currents go over to the continuum current operators
$j_0(x)$ and $j_0^5(x)$.  Obviously the local lattice charge density
operator,
whose form is fixed by the way in which one introduces gauge invariance,
cannot have this property because the normal ordered version of this
operator
satisfies the identity $ j_0(i)^3 = j_0(i)$ for all values
of the lattice spacing (since only the  charges
$0,1,-1$ can exist on a single lattice site).  On the other hand, as we
will show,
the Fourier transform $j_0(k)$ can be treated as a boson operator and the
the dynamics of the theory tells us that the current operators of the
continuum
theory are obtained by forming an appropriately regulated version of
these lattice operators.

In order to make our discussion essentially self-contained we begin by
briefly reviewing the $A_0 = 0$ gauge treatment of the Hamiltonian version
of the continuum Schwinger model.  We discuss: the need for imposing a
state condition,
such as restricting to {\it gauge-invariant states\/}; why
only the total $Q = 0$ sector of the theory can exist at finite energy;
and why different sectors of gauge-invariant states exist and are labelled
by a continuous parameter $-1/2 \ge \epsilon \le 1/2$,
which can be identified
as a {\it background electric field\/}.  Finally,
we review the Hamiltonian derivation of the fact that the electric
charge density is a free  massive Bose field and the role played
by the anomalous commutator of the electric and axial charge density
operators in the derivation of this result.  After reviewing the continuum
theory we set up and discuss the physics of the lattice version of the
Schwinger model in  $A_0(x) = 0$ gauge.  We then parallel the
continuum arguments as closely as possible for a variety of
fermion derivatives.  A careful treatment of the Hamilton equations of
motion
for the Fourier transform of the charge density operator leads to
an understanding of how regulated versions of these operators go over
to the point-split operators of the continuum theory and the sense in
which
these regulated operators can be treated as Bose fields.  The difference
between the way in which things work for generalized SLAC-type derivatives
and Wilson-type derivatives becomes clear due to this discussion, as does
the connection between the strong and weak coupling theory for generalized
SLAC-type derivatives.

\section{The Continuum Schwinger Model}

Hamiltonian formulations of the continuum Schwinger model have been
discussed in the literature\cite{Manton,Murayama}.  Our discussion will
parallel
these discussions to a degree but will differ in important details.  Our
goal
is to allow the reader to understand the important features of the
Schwinger model without unnecessary formalism.

As we have already noted, the Schwinger model is simply QED in $1+1$
dimensions, and
has a Lagrangian density given by:
\be
{\cal L} = \bar \psi ( i \partial_\mu \gamma _\mu + e A_\mu \gamma_\mu )
\psi
 -\frac{1}{4} F_{\mu\nu} F^{\mu\nu}.
\label{Lagr}
\ee

In $1+1$ dimensions there are only three anti-commuting
$\gamma$ matrices, $\gamma_0, \gamma_1, \gamma_5$, and so they can be
realized in terms of the Pauli $\sigma$-matrices:
\begin{eqnarray}
&&\gamma_0 = -i \sigma_x, \nonumber \\
&&\gamma_1 = - i \sigma_y, \nonumber \\
&&\gamma_5 =  \gamma_0 \gamma_1 =  \sigma_z.
\end{eqnarray}

In order to enable us to give the most physical treatment of
gauge-invariance
of the theory we choose to work in temporal, or $A_0(x)=0$, gauge.
Making this choice the Lagrangian density becomes:
\be
{\cal L} = \bar \psi ( i \partial_\mu \gamma _\mu - e A(x) \gamma_1 ) \psi
 +\frac{1}{2} (\partial_0 A(x))^2.
\label{Lagrtwo}
\ee
Here, for convenience, we have denoted the spatial component of the
vector potential as $A(x)$ and dropped its subscript.
Eq.(\ref{Lagrtwo}) tells us that the electric field,
\be
E(x) = \partial_0 A(x) ,
\ee
is the canonical momentum conjugate to $A(x)$ and it has the
usual equal-time commutation relations with $A(x)$:
\be
[E(x),A(x')] = -i\delta(x-x').
\label{coma}
\ee
Similarly, the fermion operators satisfy the anti-commutation relations
\be
\{ \psi^{\dagger}_\alpha(x), \psi_\beta(x') \}
= \delta(x-x') \delta_{\alpha \beta}.
\label{comp}
\ee
It follows immediately that the Hamiltonian in  $A_0 = 0$ gauge is
\be
H = \int {\rm d} x \left[ \frac {E(x)^2}{2} +
 \psi^{\dagger}(x) \left ( i \partial _1 + i e A(x) \right)
 \sigma _z \psi(x) \right].
\ee

There is an essential piece of the physics of working in $A_0 = 0$ gauge
which requires discussion.  Since we begin by setting $A_0 = 0$ in the
Lagrangian,
we cannot vary ${\cal L}$ with respect to $A_0$ or $\partial_0 A_0$,
and so we do not obtain Gauss' law
\be
 G(x) = \left(\partial _x E(x) - e \psi^{\dagger}(x) \psi(x) \right) = 0
\label{gauss}
\ee
as an operator equation of motion.  In fact, using the canonical
commutation relation, Eq.(\ref{coma}), we see that
\begin{eqnarray}
 e^{-i \int {\rm d} y \alpha(y) A(y)}\,G(x)\, e^{i \int {\rm d}y \alpha(y)
A(y)}  &=&
 \left(\partial_x ( E(x) + \alpha(x) ) - e \psi^{\dagger}(x)
\psi(x)\right) \nonumber\\
 &=& G(x) + \partial_x \alpha(x) .
\end{eqnarray}
This means that even if we start with a state $\ket{\phi}$ for which
\be
	G(x) \ket{\phi}	= 0 ,
\label{statecondition}	\ee
we can generate states of the form
\be
  \ket{\phi_{\alpha}} = e^{i \int {\rm d}\xi \alpha(\xi) A(\xi)}
\ket{\phi}
\label{generalgauss}
\ee
for which
\be
	G(x) \,\ket{\phi_{\alpha}} = \partial_x \alpha(x) \ket{\phi_{\alpha}} .
\label{irred}
\ee
Fortunately, the operators $G(x)$ (which we identify
with the generators of time-independent gauge transformations) commute
with
one another and with $H$, and so they can all be simultaneously
diagonalized.
Thus we are free to impose Eq.(\ref{statecondition}) as a state condition
because
the Hamiltonian cannot take us out of this sector of the Hilbert space.
Actually,
we are free to impose the more general condition of Eq.(\ref{irred}) for
any arbitrary
function $\alpha(x)$.  What all this means is that the canonical
quantization of the Schwinger model in $A_0 = 0$ gauge produces not one,
but rather an infinite number of theories distinguished from one another
by the fact that they have, in addition to the dynamical fermion fields,
different static classical background charge distributions
$\rho(x)_{\rm class} = \partial_x \alpha(x)$.
This shouldn't be a surprise because one should be able to
formulate QED in the presence of an arbitrary distribution of static
classical
background charges. By quantizing in $A_0 = 0$ gauge all we are doing is
obtaining all of these possibilities at the same time.

Since, on physical grounds, we are not interested in formulating the
Schwinger model
in the presence of any non-dynamical charge density, it is customary
to limit attention to the so-called {\it gauge-invariant\/} states defined
by the condition that $\rho(x)_{\rm class} = 0$.  Note that this doesn't
quite reduce us
to a single possibility since all it means is that 
$\partial_x \alpha(x) =0$
or, in other words, $\alpha(x)$ can be an arbitrary constant.  If we make
such a
transformation we shift the operators $E(x)$ by a constant, which means
that
we are free to formulate the theory in the presence of a constant
background
field $\epsilon$.
If we worked in finite volume this would amount to allowing for the
possiblity
that there are non-vanishing classical charges on the boundaries;
i.e. the remaining sectors of the
theory differ by a choice of boundary conditions.
One key question associated with the Schwinger model is whether or not the
physics
is different for different values of the background field.  In particular,
does
the ground-state energy density, which is certainly different for the free
theory, depend upon the value of $\epsilon$ when  interacting fermions
are introduced into the game.

A simple argument given by Coleman\cite{Coleman} shows that values of
$\epsilon$ which
differ by an integer must be equivalent to one another.  Before giving the
details
of the argument it is important to note that in one dimension the
solution to the equation
\be
\partial_x E(x) = \sum \limits_{j}^{} e_j \,\delta(x-x_j)
\label{discrete}
\ee
for a set of charges $e_j$ located at positions $x_j$ only has a finite
energy solution
when the total charge $\sum_j e_j = 0$.  This is so
because Eq.(\ref{discrete}) tells us that in the regions
between the  two charges
$E(x)$ is constant and it changes by an amount $e_j$ at each point
$x_j$.  If the sum of the $e_j$'s is not zero then, assuming the field
vanishes
to the left of the first charge at $x_1$, the field must continue to
infinity to the
right of the last charge.  This means that in order to minimize the
 field energy
$\int E^2(x)/2$ one or more of the charges must move off to infinity
leaving behind
a totally neutral system.  In particular, if we assume
no background field
then the energy of a pair of particles with charges $\pm 1$ separated by
a distance $s$ is $s/2$.  In the presence of a background field
$\epsilon > 0 $ the situation is different.
When the field is present, there is
a background energy density equal to
$\epsilon^2/2$.  If one now separates a pair of
charges oriented so as to reduce the field between the charges to
$\epsilon - 1$,
the total change in the energy of the system is given by
\be
\delta {\cal E} = \frac{\left( - s \epsilon^2 + s (\epsilon -1 )^2
\right)}{2}
 = \frac{s (1 - 2 \epsilon)}{2},
\label{chet}
\ee
where the term $-s \epsilon^2$ occurs because in the region
of length $s$ we have
replaced the original background field $\epsilon$ by $\epsilon -1$.
 From Eq.(\ref{chet})  it follows that for
$\epsilon < 1/2$  increasing the separation
between the charges costs energy,
while for  $\epsilon > 1/2$ separating the charges
will gain energy (i.e., by moving the charges off
to infinity one reduces the
background field to $\epsilon^{'} = \epsilon - 1$ and gains an
infinite amount of energy).
Clearly with that kind of energy gain nothing can stop this process from
happening and, since the only change
in the problem is that now there will be  pairs of
charges at $\pm \infty$, it will continue until the background field is
reduced
to the region $-1/2 \le \epsilon \le 1/2$.
For historical reasons this reduced range
of $\epsilon$ is usually
parametrized by an angle $\theta = 2\pi\epsilon$ and is
one of the two angles which label the exact solutions to the continuum
Schwinger
model\cite{Lowenstein,Coleman}.

If we work in the sector of physical states for which
\be
Q_{\rm tot} \ket{\phi} = e \int
\limits_{-\infty}^{\infty} {\rm d}\xi\, \rho(\xi) \ket{\phi} = 0,
\ee
we can solve for $E(x)$ in terms of $\rho(x)$
\be
E = e \int \limits_{}^{x} {\rm d} \xi \, \rho(\xi),~~~~Q_{\rm tot} =
e \int \limits_{-\infty}^{\infty} {\rm d} \xi \, \rho(\xi) = 0.
\ee
Substituting this into the Hamiltonian we obtain
\be
H = \int {\rm d} x  {\tilde \psi}^{\dag}(x)  i \partial _x
\sigma _z \tilde \psi(x)
- \frac {e^2}{4} \int {\rm d}x {\rm d} y \tilde \rho(x) |x-y|
\tilde \rho(y),
\ee
where $\displaystyle
\tilde \psi(x) = e^{-i\int_{-\infty}^{x} {\rm d}\xi\, A(\xi) } \psi(x)$.
This field transformation enables us to eliminate the
term $A(x)$ from the Hamiltonian
and simultaneously preserve the canonical commutation relations of
operators
$\psi(x)$, $\psi^{\dag}(x)$.  It is important to observe
that even if we had not
been able to eliminate $E(x)$ from the Hamiltonian we could have still
made this definition but it would not have been particularly useful since
in that case $E(x)$ would have non-trivial equal time commutators with
the fermion fields and we couldn't use the canonical quantization rules
to carry out computations.  Note, in what follows we will, by abuse of
notation, drop the tilde and simply write $\tilde \psi(x)$ as $\psi(x)$.

The content of the exact solution of this model is that it is
the theory of a free boson of mass $m^2 = e^2/\pi$ and, moreover,
the charge density operator $\rho(x)$ can be used as an interpolating
field for this particle because it satisfies a free field equation with
the same mass.  To see
how this happens all we need to do is derive the Heisenberg
equations of motion  for $\rho(x)$.

The time derivative of $\rho(x)$ is
\be
\partial_0 \rho(x) = \frac {1}{i}[\rho(x),H].
\ee
Since $\rho(x)$ commutes with itself, we use canonical equal
time anti-commutation
relations for  the fermionic fields Eq.(\ref{comp}) and
obtain:
\be
\partial_0 \rho(x) = \partial_x j(x),
\label{currcons}
\ee
where $j(x) = \psi^{\dag} (x) \sigma_z \psi(x)$.  Eq.(\ref{currcons})
simply states
that divergence of the vector current vanishes; i.e., the vector current
is
conserved.

The second derivative of the charge density operator is now given
by
\be
\partial_0^2 \rho(x) = \frac {1}{i}[ \partial_x j(x),H],
\ee
which evaluates to
$$
\partial_0^2 \rho(x) =  \partial_x^2 \rho(x)
- \frac {e^2}{4} \int {\rm d}y_1 {\rm d}y_2 |y_1-y_2|
\left ( \rho(y_1) [-i \partial_x j(x),\rho(y_2)]
+ [-i \partial_x j(x),\rho(y_1)]\rho(y_2) \right ).
$$

The key point in the solution of the Schwinger model is the
commutator of $j(x)$ and $\rho(x')$. It is known that
this commutator  acquires a Schwinger term
which we will compute by considering Fourier components of the currents:
\be
\rho(x) = \int \frac {{\rm d}k}{2\pi} e^{-ikx} \rho_k,~~~~
j(x) = \int \frac {{\rm d}k}{2\pi} e^{-ikx} j_k.
\ee

By introducing creation and annihilation operators for the upper $u_k$ and
lower $d_k$ components of the fermion fields with standard anticommutation
relations:
\be
\{u^{\dag}_k,u_q\} = 2\pi \delta(k-q),~~~\{d^{\dag}_k,d_q\} = 2\pi
\delta(k-q),
\ee
one obtains:
\be
[j_k,\rho_q] = \int \frac {{\rm d}l}{2\pi} \left (
u^{\dag}_{l-k}~u_{l+q} - u^{\dag}_{l-k-q}~u_{l} - ( u \to d)
\right ).
\label{commom}
\ee
At first sight, this is zero, since the integration momenta
$l$ can be shifted $l \to l - q$ in the first term of the integrand.
This, however, is not true. The problem is that the momenta shifts
can be safely done only in the operators that  are normal ordered with
respect to the vacuum state, otherwise the difference of two
infinite $c$-numbers appears. Since, in this basis, the $e=0$ Hamiltonian
$H_0$ reads:
\be
 H_0 =
\int \frac{{\rm d}k}{2\pi}\, k \,\left( u^{\dag}_k u_k - d^{\dag}_k
d_k\right),
\ee
the vacuum (the lowest energy eigenstate of $H_0$)
is obtained by filling all negative energy
states
\be
\ket{{\rm vac}} = \prod _{ k < 0} u^{\dag}_k   \prod _{ k > 0} d^{\dag}_k
\ket{0} ,
\label{ffvac}
\ee
where $\ket{0}$ is the state annihilated by the $u_k$'s and $d_k$'s.
One may see, that for $q \ne -k$ in Eq.(\ref{commom}), the right
hand side annihilates the vacuum and hence momenta shifts are
allowed. For $k = -q$, however, this is not the case, and that
can be easily seen by considering $[j_k,\rho_{-k}]\ket{ {\rm vac}}$.
One finally obtains:
\be
[j_k,\rho_q] = \frac {k}{\pi} 2\pi \delta(k+q),
\label{commomres}
\ee
which translates to:
\be
[j(x),\rho(x')] = \frac {i}{\pi}\partial_x \delta(x-x').
\label{anomcom}
\ee
Consequently
\be
[-i \partial_x j(x),\rho(x')] =  \frac {1}{\pi}\partial_x^2 \delta(x-x'),
\ee
and we obtain:
\be
\partial_0^2 \rho(x) = \partial_x^2 \rho(x)
- \frac {e^2}{2\pi}
\int {\rm d}y_1 {\rm d}y_2 |y_1-y_2| \rho(y_1) \partial_x^2 \delta(y_2-x).
\ee
Integrating  by parts twice and using
$ \partial_x^2 |x - x'| = 2\delta(x-x')$,
we obtain:
\be
\partial_0^2 \rho = \partial_x^2 \rho - \frac {e^2}{\pi} \rho.
\label{dyn}
\ee
We see therefore, that the charge density operator $\rho(t,x)$
satisfies the equation for the free field with the mass
$\mu^2 = e^2/\pi$.

Let us take another look at the role of the
anomalous commutation relation and the gauge invariance
in the exact solution of the Schwinger model. First consider
the case $e=0$. The equations of motion
\be
[\rho_k,H_0]=k j_k,~~~[j_k,H_0]=k \rho_k,
\ee
allow us to write the free fermion Hamiltonian as a
quadratic polynomial in  $\rho_k$ and $j_k$:
\be
H_0 = \frac{1}{2} \int \limits_{0}^{\infty}
 {\rm d}k \left( \rho_k \rho_{-k} +  j_k j_{-k}
                     \right ) ,
\label{hzero}
\ee
since, combined with the anomalous commutator Eq.(\ref{commomres}),
it produces exactly the same Heisenberg equations of motion.
Since the gauge invariance of the theory allowed us to eliminate
$A(x)$ from the
Hamiltonian once $E(x)$ was replaced by the Coulomb interaction
written in terms of the operators $\rho_k$ alone,
the full Hamiltonian is obtained
by adding the operator
\be
H_I =
e^2 \int \limits_{0}^{\infty} \frac {{\rm d}k}{2\pi}~
\frac {\rho_{k}\rho_{-k}}{k^2}
\label{hint}
\ee
to $H_0$. Obviously, $H_I$ is also a quadratic polynomial
in $\rho_k$ and therefore, thanks to the equations of motion, the
anomalous commutator of the spatial and temporal components of the
vector current  and the gauge invariance,
the total Hamiltonian is quadratic in $\rho_k$ and $j_k$. 
This makes the theory completely
solvable in the
continuum.  We will discuss just how much of this picture survives when
one moves
from the continuum to lattice version of the theory in the next section.

To complete the usual {\it bosonization\/} of the theory we observe that
$\rho_k$ and $j_k$ don't satisfy canonical commutation
relations, however a simple rescaling remedies this problem and at the
same time casts the Hamiltonian into a more familiar form.
To be precise, since
$\rho_k$ has no $k=0$ term\footnote{$\rho_0 \ne 0$ would imply that
the system is not neutral and that would violate the state condition
$G(x)\ket{\phi}=0$.}, we can define
\be
\sigma_k = \frac{\sqrt{\pi}}{k}\, \rho_k  ,~~~~~\Pi_k = \sqrt{\pi}\, j_k.
\label{rescaling}
\ee
Then, using Eq.(\ref{commomres}), we see that
\be
[\Pi_k,\sigma_q] = 2\pi \delta(k+q),
\ee
and the Hamiltonian takes the form
\be
H = \int_{0}^{\infty} \frac{{\rm d} k}{2 \pi} \left( \Pi_k \Pi_{-k} +
	(k^2 + \frac{e^2}{\pi} ) \sigma_k \sigma_{-k} \right).
\label{eq37}
\ee
Given the canonical commutation relations for $\Pi_k$ and $\sigma_k$ and
this form of the Hamiltonian, it is obvious that we are dealing with the
theory
of a free massive Bose field.

Let us now turn to the question of the dependence of the theory on
the background electric field, or rather to the more general question of
what happens in the Schwinger model if we introduce static classical
charges.  The remarkable property of the Schwinger model is that
independent of their magnitude these charges are screened completely.
Understanding how this occurs will fully answer the question of how
the theory depends upon a background electric field, since we already
noted that
having a background field of magnitude $-1/2 \le \epsilon \le 1/2$
corresponds
to having classical charges of magnitude $\pm \epsilon$ on the boundaries
(or equivalently at $\pm \infty$ ).

 From the solution of the theory in terms of $j(x)$ it is easy to
understand
the screening phenomena, since it follows immediately from Eq.(\ref{dyn}).
Let us consider the Schwinger model with two external charges of the
opposite
sign:
\be
\rho_{\rm ext}(x) = eQ_{\rm ext}
\left ( \delta(x-x_1) - \delta(x - x_2) \right ).
\ee
As we have seen already,
Eq.(\ref{statecondition}) gets modified to include the external charge
density.
For this reason the part of the Hamiltonian corresponding to the Coulomb
interaction
acquires an additional term and the new equation of motion becomes:
\be
\partial_0^2 \rho = \partial_x^2 \rho - \mu^2 \left ( \rho(x)
+ \rho_{\rm ext}(x) \right ),
\label{dynext}
\ee
where $\mu^2=e^2/\pi$. This equation implies that there is now a
classical,
time-independent component of the charge density operator induced
by the external charge which satisfies:
\be
\rho_{\rm ind}(x) = -\mu^2 e Q_{\rm ext}
\int \frac {{\rm d}k}{2\pi} \frac {\cos(k|x-x_1|)}{k^2 + \mu^2} -
(x_1 \to x_2).
\ee
Computing the integral, we obtain for the induced charge density:
\be
\rho_{\rm ind} = -\frac {eQ_{\rm ext} \mu}{2} \left ( e^{-\mu|x-x_1|} -
 e^{-\mu|x-x_2|} \right ),
\ee
which, as advertised, screens the external charge densities.
One interesting feature of the screening is that two external charges
get screened independently from each other \cite{Murayama}.
Note also that the screening occurs  on the scales
$\Delta x \sim 1/\mu$, which for small coupling constant
can be rather large.  Nevertheless, if we now move the external charges
off to infinity, so as to go over to the sector which in the
free theory would have an external background field, we see that
this field is totally screened in the groundstate of the interacting
theory.
Moreover, since all of the screening
takes place within a finite distance of the boundary, there is no
contribution
to the groundstate energy density coming from the background field.

We should point out that while the previous computation makes it clear
that
there shouldn't be a change in the energy density of the groundstate, it
is
not at all obvious that there isn't a finite change in the energy of the
state
due to the regions surrounding the screened external charge.  In fact,
there
clearly is such a change when the external charges are located at a finite
distance from one another; however, the question of what happens as one
moves these charges
to plus and minus infinity is a bit subtle.  The crux of the issue
has to do with a definition of the limiting process.  As will
become apparent in a moment the conventional treatment of the Schwinger
model
amounts to a prescription in which one defines the Hamiltonian of the
system
as a limit
\be
	H = \lim_{\Omega \rightarrow \infty} H_{\Omega}
	= \lim_{\Omega \rightarrow \infty} \int {\rm d}\xi\, H(\xi)
\ee
where $\Omega$ is the closed finite interval $\Omega = [ -\omega,
\omega]$.
With this definition in mind the usual prescription is to first take the
classical
background charges to plus and minus infinity and then to take the limit
$\Omega \rightarrow \infty$.  Given this prescription it is clear
that the total
Hamiltonian defined in this way never sees the classical screened charges
and
therefore there is no change in the vacuum energy.  In order to see that
this is the
usual prescription which follows from {\it bosonization\/} of the model
let us
go back to Eq.(\ref{eq37}) and  modify it
to include the possibility of having an arbitrary external classical
charge
density $\rho_{\rm ext}(x)$. In configuration space we obtain:
\be
	H = \int {\rm d}x \left( \frac{1}{2} \Pi(x)^2 + \frac{1}{2} (\partial_x
\sigma(x))^2
	                            + \frac{e^2}{\pi} (\sigma(x) + \epsilon(x)
)^2 \right),
\ee
where $\epsilon(x)$ is the function which satisfies the equation
\be
	\rho_{\rm ext}(x) = \frac{1}{\sqrt{\pi}} \partial_x \epsilon(x) .
\ee
Now, if we set $\epsilon(x)$ equal to a constant $\epsilon$ we see that
all we have to do
is define
$\tilde \sigma(x) = \sigma(x) + \epsilon$ and the Hamiltonian becomes
identical
to the one without a background field:
\be
	H = \int {\rm d}x \left( \frac{1}{2}
\tilde \Pi(x)^2 + \frac{1}{2} (\partial_x \tilde \sigma(x))^2
	               + \frac{e^2}{\pi} \tilde \sigma(x)^2 \right) .
\ee
This is the usual way of handling this issue and so we see that
this treatment says
that the groundstate energy is independent of the external
constant background field, which corresponds to the prescription we gave
above.

To complete our discussion of the continuum Schwinger model we present
another way of seeing the screening of the classical background field
which doesn't require working with the exact solution to the problem,
but only the anomalous commutation relation of $\rho(x)$ and
$j(x)$.  The key to this discussion is the introduction of the conserved
{\it gauge-dependent\/} current
\be
	\tilde{j}(x) = j(x) + \frac{e}{\pi} A(x) .
\label{jtilde}
\ee
Obviously, since $A(x)$ doesn't commute with the {\it gauge-generators\/}
$G(x)$ defined in Eq.(\ref{gauss}), this current mixes states
which satisfy different forms of the general state-condition defined in
Eq.(\ref{generalgauss}).
This means that we should think of $\tilde{j}(x)$ as operating in the full
Hilbert space of the
theory obtained by canonical quantization in $A_0 = 0 $ gauge
without imposing any gauge condition.  To show
that $\tilde{j}(x)$ is conserved we commute it with the
Hamiltonian to obtain
\be
   \partial_0 \tilde{j}(x) = \frac{1}{i} \left[ \tilde{j}, H \right]
   =  \frac{1}{i} \left[ j(x), H \right]
 + \frac{e}{i\pi} \left[ A(x), H \right].
\ee
Now, a slight rewrite of the derivation of Eq.(\ref{dyn}) gives
\be
   \frac{1}{i} \left[ j(x), H \right] =
   \partial_0 j(x) = \partial_x \rho(x)
- \frac{e^2}{\pi} \partial_x^{-1} \rho(x)
   = \partial_x \rho(x) -\frac{e}{\pi} E(x)
\ee
and since by construction $\partial_o A(x) = E(x)$, we obtain
\be
	\partial_0 \tilde{j}(x) -\partial_x \rho(x) = 0,
\ee
which means the current is conserved.  Integrating this equation over all
space
we obtain, under the usual assumptions about surface terms, that
\be
	\left[ \int {\rm d}x\,\tilde{j}(x), H \right] = 0 ,
\ee
a fact we will use in a moment.

\begin{figure}
\begin{center}
\leavevmode
\epsfxsize=5in
\epsfbox{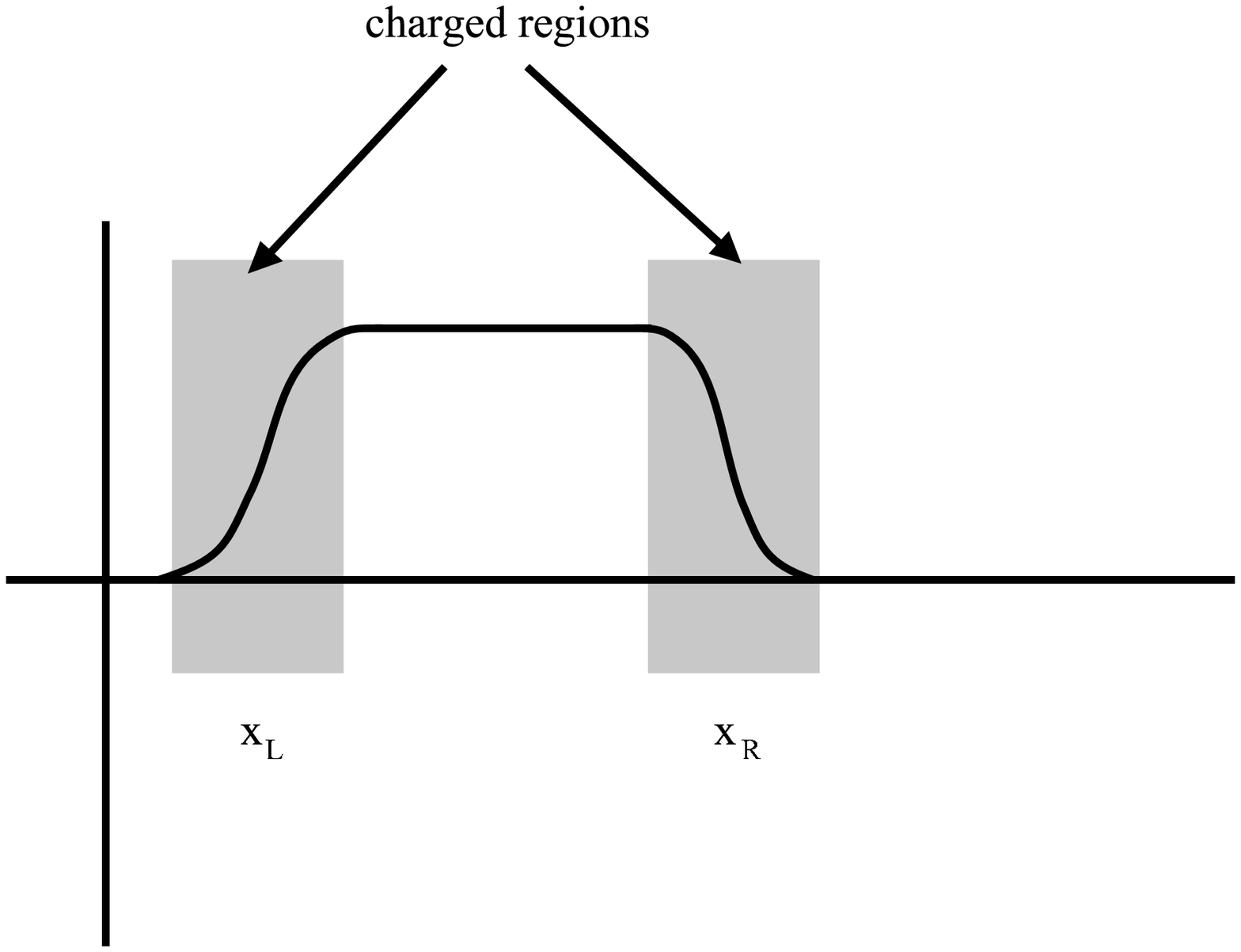}
\end{center}
\caption[alphax]{   }
\label{alphax}
\end{figure}

To understand the significance of the fact that $\tilde{j}(x)$ is
conserved
imagine that
we start in a sector of the theory whose lowest energy state satisfies
$\bra{0} G(x) \ket{0} = 0$.  Next consider the
transformed state
\be
	U(\alpha)\,\ket{0} =
	e^{i \int {\rm d}\xi \alpha(\xi) \left(j(\xi) + \frac{e}{\pi} A(\xi)
\right)} \ket{0}.
\ee
We already saw in Eq.(\ref{generalgauss})
and Eq.(\ref{irred}) that the effect of the
term proportional to $A(\xi)$ in the exponent
is to shift the field $E(x)$ so that
\be
\bra{0} U^{\dag}(\alpha)\, G(x)\, U(\alpha) \ket{0}
= \frac{e}{\pi} \partial_x \alpha(x).
\ee
This equation says
that $U(\alpha)$ takes us from a state with no background charge density
to one with background charge density equal to $ e \partial_x
\alpha(x)/\pi$.
Similarly, it follows from the commutations relations of $\rho(x)$ and
$j(x)$ and
an integration by parts, that
\be
\bra{0} U^{\dag}(\alpha)\,\rho(x)\, U(\alpha) \ket{0}
= -\frac{e}{\pi} \partial_x \alpha(x) .
\ee
Thus, the total effect of applying $U(\alpha)$ to the vacuum of
sector of the theory with no classical charges is to map this state into a
sector which has a non-vanishing
classical charge density and at the same time to produce a fermionic
charge polarization
which cancels it exactly.  Now imagine that $\alpha(x)$ is chosen as in
Fig.\ref{alphax}.
Since $\partial_x \alpha(x)$ vanishes except in the two narrow regions
around $x_L$ and $x_R$
we see that the effect of this operator is to map the original state into
one which has
equal and opposite classical charge densities around $x_L$ and $x_R$ and
induced cancelling
fermionic polarization charge densities.  As we move $x_L$ and $x_R$ to
minus and plus infinity respectively the function $\alpha(x)$ becomes a
constant and
in the limit, the fact that $U(\alpha)$ commutes with $H$ implies that
\be
	\bra{0} U^{\dag} H U \ket{0} = \bra{0} H \ket{0}.
\ee
Hence,
the energy of the vacuum of the sector with an arbitrary background field
is the same as the energy of the vacuum of the sector with no
background field, which agrees
with the previous argument for the bosonized version of the theory.

\section{The Lattice Schwinger Model}

Let us now discuss the Hamiltonian version of the Schwinger model on a
lattice.
In the Hamiltonian formalism time is continuous and space is taken to be
an
infinite lattice whose points are separated by a distance $a$.
As in the continuum, we work in $A_0 = 0$ gauge.  Furthermore, we
introduce fermionic variables $\psi^{\dag}_n$ and $\psi_n$ associated with
each site
on the spatial lattice and replace the continuous fields $A(x)$ and $E(x)$
by conjugate
variables $A_n$ and $E_n$ associated with the link $(n,n+1)$ joining the
sites $n$ and $n+1$.
This leads to a lattice Hamiltonian of the form
\be
H = H_E + H_f,
\ee
where
\be
H_E = \frac {a}{2} \sum_{n} E_n^2,~~~~
H_f = \sum_{n,n^{'}} (\psi^{\dag}_n)^{\alpha} K(n-n^{'})_{\alpha \beta}
\,e^{-ie\,\sum_{j=n}^{n^{'}-1} A_j} (\psi_{n^{'}})^{\beta}.
\ee
Here the kinetic term $K(n-n^{'})_{\alpha\beta}$ is a two-by-two
matrix for each value of
$n-n^{'}$, the fermion fields satisfy the anti-commutation relations
\be
	\left\{ (\psi^{\dag}_n)^{\alpha} , (\psi_{n^{'}})^{\beta} \right\} =
\,\delta_{n,n'}\delta_{\alpha,\beta}
\ee
and the link fields satisfy the usual harmonic oscillator commutation
relations
\be
   \left[ A_n , E_{n^{'}} \right] = i \delta_{n,n^{'}} .
\ee
Note that the fermion fields are dimensionless and in order to make the
connection to
continuum fields we will have to rescale them by a factor of $1/\sqrt{a}$
to give
them dimensions of ${\rm mass}^{1/2}$.
In direct analogy to the continuum theory, the eigenvalue of the operator
$E_n$ is
the electric flux carried by the link $(n,n+1)$.  Since, as we have seen,
the operator
$e^{-i e\,A_n}$ shifts the flux on the link $(n,n+1)$ by $e$ it follows
that
if we define the normal ordered charge density operator to be
\be
	\rho_n = :(\psi^{\dag}_n \psi_n ): ,
\ee
then the operators
\be
    G(n) = E_{n+1} - E_n - e \rho_n
\ee
commute with the Hamiltonian. Hence, similar to  the continuum,
we are free to impose the discrete version of Gauss' law
\be
	G(n)\,\ket{\phi} = \rho^{\rm class}_n \,\ket{\phi}
\label{latticegauss}
\ee
as a general state condition. Therefore we see that
the lattice and continuum versions of the Schwinger model
are essentially the same, in that canonical quantization in $A_0 = 0$
gauge
gives not one version of two-dimensional QED but rather an infinite number
of versions of the theory corresponding to quantizing in the presence
of an arbitrary classical background charge distribution.
Note that the form of Gauss' law expressed in Eq.(\ref{latticegauss})
requires
us to use the local charge density operator $\rho_n$ as the lattice analog
of the
continuum charge density operator.

Paralleling the discussion of the continuum theory as closely
as possible, we focus attention on the zero charge sector of the space of
gauge-invariant states; i.e., the ones that satisfy
the state condition
\be
	G(n) \ket{\phi} = 0 .
\ee
Once again, in this sector we can explicitly solve for $E_n$ in terms
of $\rho_n$ and eliminate the factors of $e^{i e\,A_n}$ by incorporating
them in the definition
of $\psi_n$.  In this way, in the $Q = 0$ sector of gauge-invariant
states, the lattice
Hamiltonian can be written as:
\be
H = H_f - \frac{e^2~a}{4}~\sum_{n,m} \rho_n |n-m| \rho_m.
\ee
Because the kinetic term $K(n-n^{'})_{\alpha \beta}$ is a function of the
difference of $n$ and $n^{'}$ we can write the Hamiltonian in momentum
space
as:
\be
H =  \int \limits_{-\pi/a}^{\pi/a}
\frac {{\rm d}k}{2\pi}~
\psi^{\dag}_k \left \{ Z_k \sigma_z +X_k \sigma_x \right \} \psi_k
+\frac{e^2 a^2}{4} \int \limits_{-\pi/a}^{\pi/a}\frac {{\rm d}k}{2\pi}~
\frac {\rho_{k}\,\rho_{-k}}{1-\cos ak}.
\label{momspaceham}
\ee
Here we have rewritten the Fourier
transform of $K(n-n^{'})_{\alpha \beta}$ in terms
of two functions $Z_k$ and $X_k$,
allowing for a very general class of fermion
derivatives.
Note that in Eq.(\ref{momspaceham}) and all the equations to follow we
have adopted
the convention that all momentum space operators are normalized in a way
that the continuum limit is reproduced by taking $a \to 0$ without any
additional field renormalization. For example:
\be
\{ (\psi^{\dag}_k)^{\alpha},(\psi_q)^\beta \} = 2\pi \delta(k-q)
\delta_{\alpha \beta}.
\ee

Taking our clue from the discussion of the continuum theory we now turn to
the
derivation of the Heisenberg equations of motion for the current $\rho_n$.
The first
step, namely computing
\be
	\partial_0 \rho_n = \frac{1}{i} \left[ \rho_n, H \right],
\ee
leads us to identify the result of this computation with the divergence of
the spatial component of the vector current (or, alternatively,
the time component of the axial-vector current) $j_n$.
Since the discussion to follow is necessarily a bit
detailed it is helpful to summarize what it will
show us in advance.
First, we will see that unlike the charge density operator the current
$j_n$ is intrinsically point-split as a consequence of the equations of
motion.
Second, as in the continuum, the important part
of the computation of $\partial_0 j_n$, by taking its commutator
with $H$, involves commuting the $j_n$ and $\rho_n$.
This computation will show that
one cannot solve the lattice Schwinger model exactly for
any finite value of the lattice
spacing $a$
because this lattice commutation relation is not the same as its continuum
counterpart. Note that this feature is related to the properties of the
free lattice Hamiltonian rather than being a consequence of the
interaction.
The same computation will show that the continuum
limit of the naive
commutators does not approach the continuum values for the Schwinger
model;
from this we will see why, on dynamical grounds,
one has to study what amounts to a point-split version of
$\rho_n$ in order to get the correct physics.

For the purpose of illustration, let us consider explicit forms of $X_k$
and $Z_k$
corresponding  to a number of popular fermion derivatives. In the case of
the
{\it naive\/} fermion derivative  $Z_k = \sin(ka)/a,~X_k=0$; in the case
of the Wilson fermion derivative
$Z_k = \sin(ka)/a,~~X_k = r/a~(1-\cos(ak))$; and for the SLAC derivative
one has $Z_k = k,~~~X_k=0$.
Given any one of these derivatives it is easy to find the one-particle
energy
levels of the non-interacting  Hamiltonian  $H_f$ by rotating the fields:
\be
\chi_k = U_k \psi_k,
\ee
where
\be
U_k = \e^{i \frac{\theta}{2} \sigma_y} =
\cos\left (\frac{\theta}{2} \right ) + i\sigma_y
 \sin \left ( \frac{\theta}{2} \right ),
\ee
and
$$
\cos \theta_k = \frac {Z_k}{E_k},~~~~~\sin(\theta_k) = \frac
{X_k}{E_k},~~~~~
E_k = \sqrt{X_k^2 + Z_k^2}.
$$
This unitary transformation diagonalizes the Hamiltonian:
\be
H_f = \int \limits_{-\pi/a}^{\pi/a}
\frac {{\rm d}k}{2\pi}~E_k~\chi^{\dag}_k \sigma_z\chi_k,
\ee
and if we introduce creation and annihilation operators for the
$\chi$-fields
\be
\chi_k = \left (
\begin{array}{c}
u_k \\
d_k
\end{array}
\right ),
\ee
with $\{u^{\dag}_k,u_q\} = 2\pi \delta(k-q)$ and $\{d^{\dag}_k,d_q\} =
2\pi \delta(k-q)$,
we obtain:
\be
H_f = \int \limits_{-\pi/a}^{\pi/a}
\frac {{\rm d}k}{2\pi}~E_k~\left ( u^{\dag}_k u_k - d^{\dag}_k d_k \right
).
\ee
Finally, the vacuum state of the free theory is obtained by filling the
{\it negative energy sea\/}; i.e.,
\be
\ket{{\rm vac}} = \prod _{ -\pi/a < k < \pi/a} d^{\dag}_k \ket{0}.
\ee

Given these equations it is a straightforward matter to compute the
commutator of $H$ with $\rho_n$ to obtain $\partial_0 \rho_n$:
\be
\partial_0 \rho_n = \frac {1}{i} [\rho_n, H],
\label{vcons}
\ee
which in the continuum theory is equal to $\partial_x j(x)$ (where $j(x)$
is identified
as the spatial component of the vector current, or the time component
of the axial-vector current).  Computing the commutator of $H$ with
$\rho_n$
is straightforward but we must say a few words about how we identify
$j_n$.  Basically, in order to maintain the parallel to the continuum
discussion we
define the quantity equal to $\partial_0 \rho_n$ as the lattice derivative
of $j_n$; i.e.,
\be
\partial_0 \rho_n = \frac {1}{a} \left ( j_{n+1} - j_n \right ).
\label{jdef}
\ee
With this identification, the algebra of $\gamma$ matrices in two
dimensions
ensures that the spatial component of the vector current
coincides with the temporal component of the axial
current, and therefore all the currents we are going to work
with appear to be defined. Clearly, different lattice
fermion derivatives will produce different definitions of the
spatial component of the vector current operator, an inescapable
consequence
of the Heisenberg equations of motion.

To derive an explicit form for $j_n$, we
Fourier transform Eq.(\ref{jdef}).  Defining
\be
\rho_k  =\sum_{n} \rho_n e^{ikan},
\ee
we obtain:
\be
\partial_0 \rho_k = \frac {1}{i} [\rho_k, H].
\label{momsp}
\ee
Writing the right hand side of this equation as:
\be
\frac {1}{a} \sum_n \left ( j_{n+1} - j_{n} \right ) e^{ikan} =
\frac {-2i\sin(ak/2)e^{-ika/2}}{a}j_k,
\ee
defines the Fourier transform of the spatial component
of the vector current. Explicit computation of $\rho_k$ yields:
\be
\rho_k = \int \limits_{-\pi/a}^{\pi/a} \frac {{\rm d}k_1}{2\pi}
\frac {{\rm d}k_2}{2\pi} \psi^{\dag}_{k_1} \psi_{k_2} 2\pi \delta^{\rm
lat}
(k_1 + k - k_2),
\ee
where $\delta^{\rm lat}(q)$ is the lattice $\delta$-function which
implies the momentum conservation modulo $2\pi/a$.  Focusing, for the sake
of
definiteness, on momenta $k > 0$, one finds:
\be
\rho_k = \int \limits_{-\pi/a}^{\pi/a-k} \frac {{\rm d}k_1}{2\pi}
\psi^{\dag}_{k_1} \psi_{k_1+k} +
\int \limits_{\pi/a-k}^{\pi/a} \frac {{\rm d}k_1}{2\pi}
\psi^{\dag}_{k_1} \psi_{k_1+k-2\pi/a}.
\ee

It is now a straightforward matter to compute the spatial component of the
vector
current using Eq.(\ref{momsp}):
\be
j_k = \frac {ae^{ika/2}}{2\sin(ak/2)} \left [
\int \limits_{-\pi/a}^{\pi/a-k} \frac {{\rm d}k_1}{2\pi}
\psi^{\dag}_{k_1} M(k_1,k) \psi_{k_1 + k}.
+
\int \limits_{\pi/a-k}^{\pi/a} \frac {{\rm d}k_1}{2\pi}
\psi^{\dag}_{k_1} M(k_1,k)  \psi_{k_1 + k-2\pi/a}.
\right ],
\ee
where
\be
	M(k_1,k) = \left \{ \left (Z_{k+k_1} - Z_{k_1} \right ) \sigma_z +
	 \left ( X_{k+k_1} - X_{k_1} \right ) \sigma_x \right \}
\ee
and we have used the fact that $Z_k$ and $X_k$ are periodic functions
with the period $2\pi/a$.

 From the continuum solution of the Schwinger model it is clear that we
should focus
on the Schwinger term appearing in the  commutator $[j_k^{\dag},\rho_q]$,
since it is the source of the anomalous Heisenberg equation of motion and
the reason for the mass of the photon being non-zero.  As we saw in the
previous section it suffices to take the vacuum expectation value
$\langle {\rm vac}| [j_k^{\dag},\rho_q] \ket{{\rm vac}}$
in order to compute the Schwinger term.
Direct computation yields the  following result:
\be
\langle {\rm vac}| [j_k^{\dag},\rho_q] \ket{{\rm vac}}
= 2\pi \delta(k-q)~W(k),
\ee
where the function $W$ is:
\be
W = \frac {ae^{-ika/2}}{2\sin(ak/2)} \left [
\int  \limits_{-\pi/a}^{\pi/a} \frac {{\rm d}k_1}{2\pi}
\left ( 2 Z_{k_1} - Z_{k_1 - k} - Z_{k_1+k} \right ) \cos \theta_{k_1}
+
\left ( 2 X_{k_1} - X_{k_1 - k} - X_{k_1+k} \right ) \sin \theta_{k_1}
\right ].
\label{E55}
\ee

To compare the result of this computation with the continuum result
we take the limit $a \to 0$,  in which case Eq.(\ref{E55}) simplifies and
one obtains:
$$
\lim_{ak \to 0} W = \frac {k}{\pi} \left [
\int \limits_{0}^{\pi} {\rm d}\xi \left (
\frac {{\rm d}^2 Z_\xi}{{\rm d} \xi ^2}
\cos(\theta_\xi) +
\frac {{\rm d}^2 X_\xi}{{\rm d} \xi ^2}
\sin(\theta_\xi)
\right )
\right ].
\label{anomcommw}
$$
This equation gives the $a\to 0$ limit of the anomalous commutator
for a general lattice fermion Hamiltonian and is therefore
useful for the analysis of the continuum limit of the various
choices for the fermion derivative. To get a feeling for how things work
let us consider several specific examples.

Let us begin with the case of the {\it naive\/} lattice fermion
derivative, where
$Z_\xi = \sin \xi $, $X_\xi = 0$, $E_\xi = |\sin \xi|$.
In this case we obtain:
\be
\lim_{ak \to 0} W = -\frac {k}{\pi}
\int \limits_{0}^{\pi} {\rm d}\xi \sin \xi = - \frac {2k}{\pi}.
\ee
This shows that the anomalous commutator is two times larger
than the continuum result, which implies that in the $a \to 0$ limit
the mass of the photon is two times larger than in the continuum theory.
In principle, this result should have been expected since the lattice
theory
with the naive fermion derivative
has an exact $SU(2)$ symmetry for all values of $a$
and as a consequence of this symmetry the fermion spectrum is doubled
as is evident from the form of $E_k$.
Thus, it follows that the continuum limit of the naive theory
is not the original Schwinger
model, but rather an $SU(2)$-Schwinger model which is known to have a
photon mass
which is $ 2 e^2/\pi$.

In case of Wilson fermions we have $Z_\xi = \sin \xi$,
$X_\xi = r (1 - \cos \xi)$ and $E_\xi = \sqrt{ Z_\xi^2 + X_\xi^2}$.
By explicit calculation one finds:
\be
\lim_{ak \to 0} W = -\frac {2k}{\pi},
\ee
Once again the result is two times larger than the continuum
one\footnote{The fact that the anomalous commutator for Wilson fermions
is $r$-independent in the $a \to 0$ limit
is a bit of a miracle and we do not quite
understand the reason for that.  Note however, that a similar
situation has been observed in  earlier calculations of the
chiral anomaly in the lattice Schwinger model with Wilson fermions
\cite{Ambjorn}. It is generally accepted that the continuum limit
of the Schwinger model with Wilson fermions gives correct anomaly
and correctly reproduces all other continuum results.
However, there is no direct contradiction with our statement
since, as we explained and as our result seem to illustrate,
different currents lead to different results.},
but in this case the low energy spectrum is clearly undoubled and the
reason for the discrepancy between the lattice and continuum results
must be different.

In order to clarify the underlying physics, it is
instructive to consider somewhat unconventional fermion derivatives.
Let us begin by considering a modified SLAC fermion
derivative\cite{Nason:1985yp}.
Consider the free fermion Hamiltonian defined by ($k > 0$,~$Z_{-k} =
-Z_{k}$):
\be
Z_k = k \theta \left ( \mu \frac {\pi}{a} - k \right )
 + \frac {\mu}{\mu -1 } \left (\frac {\pi}{a} - k \right )
  \theta \left ( k - \mu \frac{\pi}{a}  \right ),~~~~
X_k = 0,
\ee
and $E_k$ is  equal to $|Z_k|$.
A plot of $E_k$ is shown in Fig.2.
Computing the second derivative of $Z_k$ one obtains:
\be
-\frac {{\rm d}^2 Z_\xi}{{\rm d} \xi ^2} =
\delta \left ( \mu \frac {\pi}{a} -\xi \right )
+ \frac {\mu}{1-\mu} \delta \left ( \xi - \mu \frac {\pi}{a} \right ),
\ee
and the anomalous commutator becomes:
\be
\lim_{ak \to 0} W = \frac {-k}{\pi}
\left ( 1 + \frac {\mu}{1-\mu} \right ).
\ee

To understand the information encoded in this form of the anomalous
commutator let us
consider what happens in the continuum limit.
It is clear from the plot of $E_k$ for this modified SLAC derivative
that two species
of fermions survive in
the limit $a \to 0$.  Note however that ${\rm d}E_k/{\rm d}k$
is quite different for the two linear regions of the spectrum, which means
that the two species
propagate with very different speeds.  The anomalous commutator
is really the sum of two contributions: one coming
from $0< k < \mu \pi/a$ and the other from
$\mu \pi/a < k < \pi/a$ and these contributions can be easily identified
with the different fermions. Since both fermions are charged, they both
contribute to the anomalous commutator and to the mass gap.

Given the simple nature of this fermion derivative it is clear
how to separate the  contributions of the two fermion species
to the total current. The easiest way to do this is to put a sharp
momentum cut-off somewhere below and above the turning point $\mu \pi/a$.
With this prescription we write the charge density operator as a sum of
three contributions:
\be
\rho_k = \rho_k^{(1)} + \rho_k^{(2)} + \rho_k^{(3)},
\ee
where
\be
\rho_k^{(1)} = \int \limits_{-\mu_1 \pi/a}^{\mu_1 \pi/a}
\frac {{\rm d}k_1}{2\pi} \psi^{\dag}_{k_1} \psi_{k_1+k},~~~
\rho_k^{(3)} = \int \limits_{|k_1| > \mu_2 \pi/a}
\frac {{\rm d}k_1}{2\pi} \psi^{\dag}_{k_1} \psi_{k_1+k} +
\int \limits_{\pi/a-k}^{\pi/a} \frac {{\rm d}k_1}{2\pi}
\psi^{\dag}_{k_1} \psi_{k_1+k-2\pi/a},
\ee
and $\mu_1 < \mu < \mu_2$.  The $\rho_k^{(2)}$ provides for the
remaining contribution to the charge density and it is only sensitive to
fermions
with momenta $\mu_1\pi/a < |k| < \mu_2 \pi/a$.
Now, following our previous argument, we define the corresponding spatial
components of the vector
current by explicitly commuting the above charge densities with the
Hamiltonian.

A straightforward computation shows that in the limit of vanishingly small
lattice
spacing the anomalous commutators of the above currents are given by:
\ba
&&
[\left ( j^{(1)}_k \right )^{\dag},\rho^{(1)}_q] = -\frac {k}{\pi} 2\pi
\delta(k-q),
\nonumber \\
&&[\left ( j^{(2)}_k \right )^{\dag},\rho^{(2)}_q] = 0,
\nonumber \\
&&
[\left ( j^{(3)}_k \right )^{\dag},\rho^{(3)}_q] =
-c \frac {k}{\pi} 2\pi \delta(k-q),
\label{eqmotion}
\ea
where we introduced $c = \mu/(1-\mu)$, which is the velocity
of the fermions in the region $k \sim \pi$.
Note that in all of these formulas there are also non-vanishing
normal-ordered operators coming from large momentum excitations
which we have not displayed.
These operators annihilate the vacuum and one can argue that they are
unimportant for small $k$ physics.  This final point, which is intimately
related
to the sense in which $j_k$ can be treated as a boson operator, merits
elaboration
and we will return to it immediately after completing our discussion of
the equations
of motion.

Proceeding with our computation of the equations of motion for
$\rho_k^{(1)}$
and $\rho_k^{(3)}$ we obtain:
\ba
&&-\partial_0^2 \rho_k^{(1)} = k^2 \rho_k^{(1)} +
\frac {e^2}{\pi} \rho_k^{\rm tot},
\nonumber \\
&&-\partial_0^2 \rho_k^{(3)} = c^2 k^2 \rho_k^{(3)} +
c \frac {e^2}{\pi} \rho_k^{\rm tot},
\ea
where the total charge density operator appears on the right
hand side of these equations and so $\rho_k^{(2)}$ is still included.
Consistent with the point made above and subject to the discussion to
follow
we will set it to zero, since for small energies
fermions with such high momenta
are not excited.

 From the equations for the commutators we see that the
field $\rho_k^{(3)}$ is not canonically normalized and so we
introduce the new field $\tilde \rho_k^{3} = 1/\sqrt{c} \rho_k^{(3)}$
and its commutation relation with the corresponding current becomes
canonical. The equations of motion  become:
\ba
&&-\partial_0^2 \rho_k^{(1)} = k^2 \rho_k^{(1)} +
\frac {e^2}{\pi} \left ( \rho_k^{(1)}
+ \sqrt{c} \tilde \rho_k^{(3)} \right ),
\nonumber \\
&&-\partial_0^2 \tilde \rho_k^{(3)} =  c^2 k^2 \tilde \rho_k^{(3)} +
 \frac {e^2}{\pi} \left ( \sqrt{c} \rho_k^{(1)}+c\tilde \rho_k^{(3)}
\right ).
\label{latteqsofmotion}
\ea
The energies of elementary excitations can  be determined
from the eigenvalues of the matrix
\be
{\cal M} = \left (
\begin{array}{cc}
k^2 + e^2/\pi & \sqrt{c}~e^2/\pi \\
\sqrt{c}~e^2/\pi  & c^2 k^2 + c~e^2/\pi
\end{array}
\right ).
\label{gbmatrix}
\ee
The matrix is easily analyzed in the limit of large $c$ which corresponds
to the large slope of the fermion derivative in the region $k \sim \pi$.
Note that the original SLAC fermion derivative corresponds to
$c \to \infty$ limit.

It is then easy to see that there are two different limits in this
equation.
For small momenta $ ck^2 \ll e^2/\pi$, there are two eigenvalues:
$$
E_1 = \sqrt{c} k,~~~~E_2 = \sqrt{1+c} \frac {e}{\pi}.
$$
Hence there is a zero mass eigenstate which is
the Goldstone boson of the theory. The other excitation is the massive
one. If we consider the limit $c \to \infty$, the region of momenta
sensitive to the Goldstone mode shrinks to zero (see Fig.3)
and the mass  of the other excitation goes to infinity.

For larger momenta, $ ck^2 \gg e^2/\pi$, the mixing of two states
becomes small, they propagate independently and  their energies are given
by:
\be
E_1 = \sqrt{k^2 + \frac {e^2}{\pi}},~~~
E_2 = \sqrt{c^2 k^2 + c \frac {e^2}{\pi}}.
\ee
In this momentum region the energy of the lower excitation
approaches the result of the continuum theory of a bosonic
field with the mass $e^2/\pi$, the other excitation becomes
infinitely heavy and decouples explicitly (see Ref.\cite{Nason:1985yp}).
Hence, as we approach the limit $c \to \infty$ (which is
equivalent to the orginal form of the SLAC derivative), the continuum
limit of the theory has a massive boson of mass $m^2 = e^2/\pi$ and an
isolated state at $k = 0$ which can be identified with
a {\it seized Goldstone mode\/}\cite{KogutandSusskind}.

Since the main purpose of this paper is to provide an analytic framework
for CORE computations to follow we should point out that the fact that
there
is a Goldstone mode when $e^2/\pi \gg c k^2 $ is quite significant since
it relates to the whole question of how the strong-coupling limit of the
model connects to the weak coupling theory and whether one can get the
correct
physics by projecting onto the sector spanned by the $ e \to \infty $
eigenstates.
We will have more to say about this point in the conclusions, but first we
should
complete our discussion of terms we ignored in the commutator of $[ j_k,
\rho_q ]$.

While this preceding argument leading to Eq.(\ref{latteqsofmotion}) makes
the
lattice discussion look remarkably like the continuum Schwinger model, we
are not really
finished.  The issue which still needs discussion relates to the
interpretation
of $\rho_k$ and $j_k$ as boson fields.  This is more than an academic
issue.
Although, as
we have shown, the vacuum expectation value of the commutator of
$j_k$ and $\rho_q$ gives the required Schwinger term, computation of the
full
commutator contains an extra piece which, if it has
non-vanishing matrix elements
between states whose energy remains finite in the $a \to 0$ limit, ruins
the
interpretation of $j_k$ and $\rho_q$ as boson fields.  This is the issue
which
we will now address.

The commutation relation for $j_k$ and $\rho_q$, which is valid for
arbitrary lattice spacing, reads:
\be
[\left ( j^{(1)}_k \right )^{\dag},\rho^{(1)}_q] = \frac
{e^{ika/2}ak}{2\sin(ka/2)}
\left (
-2\pi \delta(k-q) \frac {k}{\pi} + O(q,k) \right ),
\ee
where the normal ordered operator $O(k,q)$ has the form:
\be
O(k,q) =  \theta(k-q) \left \{\int \limits_{\mu_1 \pi/a-k}^{\mu_1 \pi/a}
\frac {{\rm d}k_1}{2\pi} :\psi^{\dag}_{k_1+k} \sigma_z \psi_{k_1+q}:
+ \int \limits_{-\mu_1 \pi/a-q}^{-\mu_1 \pi/a}
\frac {{\rm d}k_1}{2\pi} :\psi^{\dag}_{k_1+k} \sigma_z \psi_{k_1+q}:
\right \}
+(k \leftrightarrow q).
\ee

We will now argue that even though the term $O(k,q)$ is not
explicitly suppressed by a power of $a$, nevertheless this operator does
not
contribute to the dynamics of any state whose energy remains finite as
$a \to 0$;
in particular, any state which  can be created by applying arbitrary
powers
of $\rho_k$ to the groundstate of the theory.
As we pointed out in the Introduction,
the operator $\rho_n$ cannot be considered a boson operator since
$\rho_n^3 = \rho_n$,
thus arbitrary powers of $\rho_n$ can produce at most three linearly
independent states
when they are applied to the groundstate.  The situation is quite
different for $\rho^{(1)}_k$
and $\rho^{(3)}_k$ which are sums of $\rho_n$'s and therefore, for an
infinite lattice,
will not satisfy an identity of this type.

The argument begins by considering the non-interacting theory and
thinking of $O(k,q)$ as acting on
a Hilbert space constructed by applying polynomials in
the current operators to the groundstate of the free theory.  For values
of
$k$ and $q$ which are small compared to $\pi/a$, the operator $O(k,q)$ can
only
act on the part of a state  which contains
left and right moving fermions with  momenta
$k \approx \pm \mu_1 \pi/a$ since it has to first absorb a high momentum
fermion and
create another one with a momentum which differs by a small amount.
This means that in order to have a matrix element of this operator between
states
generated by polynomials in $\rho^{(1)}_k$ and $\rho^{(3)}_k$ these states
have
to have non-vanishing components having fermions  with high momenta.
Thus, we have to ask how such components can be generated?

Both $\rho^{(1)}_k$ and $\rho^{(3)}_k$ are bilinears
in fermion creation and annihilation
operators and, being normal ordered,
can only absorb a fermion at one momentum and create
a replacement at another momentum.  Generically these operators are of the
general form:
\be
	\rho^{(i)}_k = \int \frac{{\rm d}k_1}{2\pi} \left( u^{\dag}_{k_1} u_{k_1
+ k}
			+  d^{\dag}_{k_1} d_{k_1 + k} \right)
\ee
and since
for the modified SLAC derivative $X_k = 0$,
the vacuum, as in the continuum, is given by Eq.(\ref{ffvac}).
If we now, for the sake of definiteness, consider
\be
	\rho^{(1)}_{k > 0} \ket{{\rm vac}} = \rho^{(1)}_k\,
	\prod _{ k < 0} u^{\dag}_k   \prod _{ k > 0} d^{\dag}_k \ket{0}
\ee
we see that almost all terms in $\rho^{(i)}_k$ annihilate the vacuum
state.  The only
terms which act non-trivially are ones were either $u_{k_1 + k}$ or
$d_{k_1 + k}$ can
absorb a particle and then either $u_{k_1}^{\dag}$ or $d_{k_1}^{\dag}$ can
create a
particle.  Clearly, for small $k > 0$
only the $d_{k_1}$ terms can act, since if $k_1 < 0 $ and $ k_1 + k > 0$
then
$d_{k_1 + k}$
can absorb a $d$ from the vacuum state and $d^{\dag}_{k_1}$ can create a
$d$.
The $u_k$ terms cannot act non-trivially because in order for $u_{k_1 +
k}$
to absorb a particle,
$k_1 + k$ has to be less than zero, in which case $k_1 < 0$ and therefore
$u^{\dag}_{k_1}$ annihilates the resulting state.  If, however, $k < 0$
then it is the
$u^{\dag}_{k_1} u_{k_1+k}$ term which acts non-trivially and
the corresponding $d$ term
annihilates the state.  In either event
the important point is that the $\rho^{(i)}_k$ only
creates and absorbs particles from the vacuum
which are within a distance $|k|$ of the
top of the negative energy sea (i.e., the fermi-surface) thus creating a
particle anti-particle
pair.

The next step is to see what happens if we apply $\rho_k^{(i)}$
to the state we just generated.
What we get is
\be
   {\rho^{(i)}_k}^2\ket{{\rm vac}}
 = \frac{1}{(2\pi)^2}
  \int {\rm d}k_1 \,{\rm d}k_2\, \left( u^{\dag}_{k_2} u_{k_2+k}
   + d^{\dag}_{k_2} d_{k_2+k} \right)\, \left( u^{\dag}_{k_1} u_{k_1+k}
   + d^{\dag}_{k_1} d_{k_1+k} \right) \ket{{\rm vac}} .
\ee
It should be clear that for almost all $k_1$ and $k_2$ in
the allowed region ${\rho_k^{(i)}}^2$
creates two low momentum particle anti-particle pairs and in fact for
given allowed
$k_1$ and $k_2$ there are $2!$ ways of getting the same two-pair state;
however, for a given
$k_1$ there is exactly one value of $k_2$ for which one can create a
higher energy one-pair
state by absorbing one of the particles in the pair created by the first
application of
$\rho^{(i)}_k$ and promoting it to higher momentum.  From this it follows
that the factor
needed to normalize this state is greater than $1/\sqrt{2!}$.  Similarly,
if one hits this
state with another power of $\rho^{(i)}_k$ almost all of the terms would
create three low energy
particle anti-particle pairs and each of these three pair states would be
created $3!$ times.
As in the previous case however there would be a single term which could
promote the
previous single higher energy one-pair state to yet higher energy.  Note,
however, that
since the normalization of this state would have to be larger than
$1/\sqrt{3!}$ (which we
are beginning to recognize as the normalization factor which goes with a
three boson state)
the coefficient of this higher energy single-pair state appearing in the
normalized
version of the state created by ${\rho^{(i)}}^3_k$ is getting smaller each
time.
If one now imagines carrying out this process $p$-times the argument
generalizes in the
obvious way.  The state obtained by applying $p$ powers
of $\rho^{(i)}_k$ to $\ket{{\rm vac}}$
is going to be mostly made of $p$ different low-energy
particle anti-particle states, each of which
will be arrived at in $p!$ ways.  Furthermore, there will be  a single
particle anti-particle pair state with individual momenta $p$ times larger
than $k$.
Since now the normalization factor of this state is bigger than $p!$,
the coefficient
of this single higher energy pair state is getting very small relative to
the
part of the wavefunction made of
 $p$ low energy pair states.  A more careful discussion
of this point would also take into account the fact that the same
procedure will
generate two-pair, three-pair, etc., parts of the wavefunction.  However,
the point
is that if we keep $k \le \Lambda$, where $\Lambda$ is a maximal
energy we wish to consider and $\Lambda a \to 0$ in the continuum limit,
then in order
to achieve the fermionic level with momenta $\mu_1 \pi/a$,
one should create a state $j_k^N\ket{{\rm vac}}$ with
\be
N \sim \frac {\mu_1 \pi}{ak} \sim\frac {\mu_1 \pi}{a\Lambda} \to \infty.
\ee
The energy of this bosonic state is
$\sim E_{\rm typ} \sim \mu_1 \pi/a \to \infty$;
and it is easy to see that the probability of finding a single high
momentum
pair state equals to  $1/N! \to 0$.
The factors of $p!$ which appear in the normalization
of the $j^p\ket{{\rm vac}}$
states thus produce the explanation of both why we can think of
$\rho^{(i)}_k$ as
a boson
operator and why $O(k,q)$ has no significant
matrix elements between normalized
states generated by applying arbitrary
powers of $\rho^{(i)}_k$ to $\ket{{\rm vac}}$.


The main point of the above discussion is that as we approach
the continuum limit
the essential physics of the model is taking place near the
top of the negative energy
sea and so it is useful to limit our attention to modified
operators $\rho^{(i)}_k$
that only have support in these regions.  For the model based upon a
modified
SLAC derivative we saw that, since the low energy spectrum
of the theory was explicitly doubled, the non-split fermion current
really was made up of two parts: the first, coming from states near
$k \sim 0$ and the other from $k \sim \pi$. Leaving aside the
complications
related to the  existence of the Goldstone mode, we see that the dynamics
of the
theory tells us that the current constructed out of the fermionic fields
with
small momenta is essentially the current with the correct continuum limit.
The large momentum part of the current decouples from the continuum limit
after the limit $c \to \infty$ is taken.  From this point of view, we see
that
the dynamics of the lattice model tells us that in order to take
the continuum limit of the
theory we have to restrict attention to only a part of the unregulated
lattice
current.  This is essentially equivalent to adopting a
point-splitting procedure
for defining the current in the continuum theory.

Though these peculiarities have been made obvious because of the explicit
doubling, our calculation of the anomalous commutation relation shows that
for the
non-point-split currents the large momentum modes do not
decouple automatically, even without fermion doubling.  To have explicit
decoupling one has to construct the currents by explicitly cutting off
the region of large momentum. If in the small momentum region the fermion
derivative is sufficiently {\it continuum-like\/} (i.e., linear) then we
are assured that:
the current constructed in this way will have correct anomalous
commutation relations modulo  corrections suppressed by inverse
cut-off; the equations of motion for this current will be free
equations of motion with an additional source term given by
high momentum fermionic modes; if the cut-off is sufficiently large in
{\it physical} energy units (as opposed to lattice units), such a current
operator
will correspond  to a continuum bosonic degree of freedom for all
low-energy
purposes.

To conclude this section let us consider a simple example of what we will
refer to as a {\it perfect Wilson model\/} for the fermion derivative.
It is defined by
$$
Z_k = k \theta \left (\frac {\pi}{2a} - k \right ) + \frac {\pi}{2a}
\sin \left ( \pi - ka \right ) \theta \left ( k - \frac {\pi}{2a} \right
),
~~~~  X_k = \frac {\pi}{2a}
\cos \left ( \pi - ka \right ) \theta \left ( k - \frac {\pi}{2a} \right
),
$$
where we have defined $Z_k$ and $X_k$ for $k > 0$, and assumed that
$Z_{-k} = Z_k$ and $X_{-k} =X_k$. The one-particle energy spectrum
is shown in Fig.4.

One may easily check (using our general result for the anomalous
commutator)
that for this model, in the limit $a \to 0$, the commutator for the
non-split
currents $[j_k,\rho_q]$
differs from the continuum limit, even though in this case
there is no doubling and the theory remains {\it continuum-like\/} up to
momenta $k \sim \pi/(2a)$.  Note, however, that if we construct a low
energy
current by restricting to the linear region of the derivative function, as
in the case of the modified SLAC derivative, we can guarantee that the
high momenta modes
do not have an influence on the dynamics of the low energy current and we
can
verify that this low-energy current and its time derivative satisfy the
desired
anomalous commutation relations.  Once we establish this fact we can
proceed to
derive the Heisenberg equations of motion for this low-energy (or
regulated)
version of the current and make the connection to the continuum theory.
Conceptually, our example of the {\it perfect Wilson\/} fermion derivative
is
very close to Wilson's original proposal.  All we have done is to enlarge
the
region of momentum space in which the lattice derivative looks identical
to the continuum derivative so as to make it easier to see why this type
of
fermion derivative works once the proper current operators have been
identified.

\section{Conclusions}

As we have said in the
Introduction our aim is to use the lattice Schwinger model
to test the idea that one can use CORE methods to map gauge theories into
highly frustrated spin  antiferromagnerst and then use the same
methods to study these spin systems.  The Schwinger model is a very good
place to test this notion since the continuum model exhibits a rich
spectrum of physical phenomena, anomalous commutators, background electric
fields,
charge screening, etc., and so it is important that any numerical
treatment of this model be able to see these effects.
We had several major goals in this paper: first, to get some analytic
control over the physics of the lattice Schwinger model in order to
understand which features of the continuum theory we might expect to
emerge
easily from a numerical computation and which might be difficult to
obtain; second,
to gain a feeling for how much of this physics we might hope to see if we
first use
CORE to map the lattice Schwinger model into a highly-frustrated
generalized
antiferromagnet and then to analyze the physics of that spin system before
carrying out detailed numerical computations; third, to get a better
understanding
of how the low-energy physics of the lattice system depends upon the
choice
of fermion derivative and why, on dynamical grounds, the lattice currents
of
interest are those which correspond to continuum point-split currents.

To accomplish our goals we studied Hamiltonian formulations of both the
continuum and lattice Schwinger model and then, by paralleling the
solution
of the continuum version of the theory in the lattice framework,
identified
those features of the lattice theory which differ from the continuum
theory
and identified the operators of the lattice theory which go over smoothly
to their continuum counterparts.  It became apparent from the treatment
of the equations of motion for the various forms of the charge density
in the lattice theory that getting the right behavior involves showing
that one is close enough to the continuum limit so that the appropriately
defined currents act as bosons;  in other words, it is not sufficient to
only show that there is a gap between the vacuum state
and the first excited state and that it numerically appears to be of the
order of $e/\sqrt{\pi}$.  At a minimum one should be able to show that the
operators $O(k,q)$ have negligible matrix elements between the computed
low-lying states of the theory.

A surprising outcome of this work was the fact that almost any fermion
derivative works for the study of the Schwinger model.  As we have seen,
the $c \to \infty$ limit of the chirality conserving modified
SLAC derivative and the perfect Wilson derivative had essentially the
same low-energy behavior.  The interesting fact was that the dynamics
of the system, while different for the two cases, managed to automatically
eliminate spurious degrees of freedom.  Basically this says that
we can use any short-range derivative, either chirality preserving
or chirality violating, to carry out numerical studies of the lattice
Schwinger model and by comparing them get additional control over how well
the numerical methods can be expected to converge.

Finally, and most pertinent to our eventual goal, is the fact that the
discussion of the modified SLAC derivative shows that the trick of using
CORE to map the system into a frustrated generalized antiferromagnet will
preserve the relevant low energy physics. The reason for this is that the
CORE method is based on defining the set of {\it retained states\/} to be
those
states which have zero energy in the limit $e \to \infty$.  This, of
course,
requires that
for these states the Coulomb term vanishes.  In other words, these
are the states for which the normal ordered charge density operator
$\rho_n$ is zero identically.  (This set of
states is generated by selecting from the four possible states per site,
only the two states having zero charge and then taking tensor products of
all
of these states.)  Note that for large $e^2$ these states are all
degenerate
to order $e^2$ and this degeneracy is lifted by the kinetic term
which acts on them by creating a pair of separated charges
and then acting a second time to bring them back to a neutral state.
A second order degenerate perturbation theory calculation shows that the
low
energy theory in the
large $e^2$ limit is that of a Heisenberg anti-ferromagnet,
which means that in this limit the theory is that of a massless particle.
Going back to Eq.(\ref{gbmatrix}) we see that perturbing in the kinetic
term
is the same as taking $e^2/\pi$ to be much greater than $k^2$ and $c k^2$.
But this
is exactly the situation in which we have one massive and one massless
mode in the
theory and the low energy physics is that of a massless boson. This
matching
of the two results at large $e^2$ would imply that the space of {\it
retained states\/}
must have a non-vanishing overlap with the true low lying states of the
theory
for finite values of $e^2$ which is all that is needed to show that the
CORE
method must work.


\newpage

\epsfverbosetrue

\begin{figure*}
\begin{center}
    \leavevmode
    \epsfxsize=10.cm
    \epsffile[58 57 401 293 ]{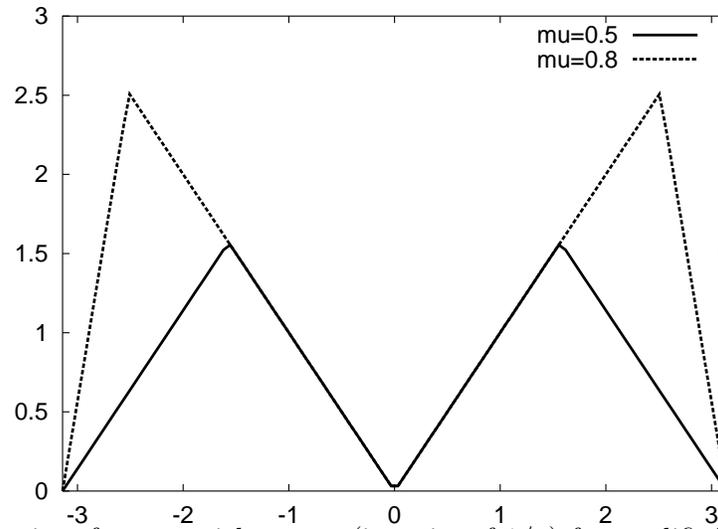}
    \hfill
   \parbox{14.cm}{
    \caption[]{The energies of one-particle states  (in units of $1/a$)
 for modified SLAC derivative for two different values of $\mu$.
}}
  \end{center}
\label{slacmod}
\end{figure*}

\begin{figure*}
\begin{center}
    \leavevmode
    \epsfxsize=10.cm
    \epsffile[58 57 401 293 ]{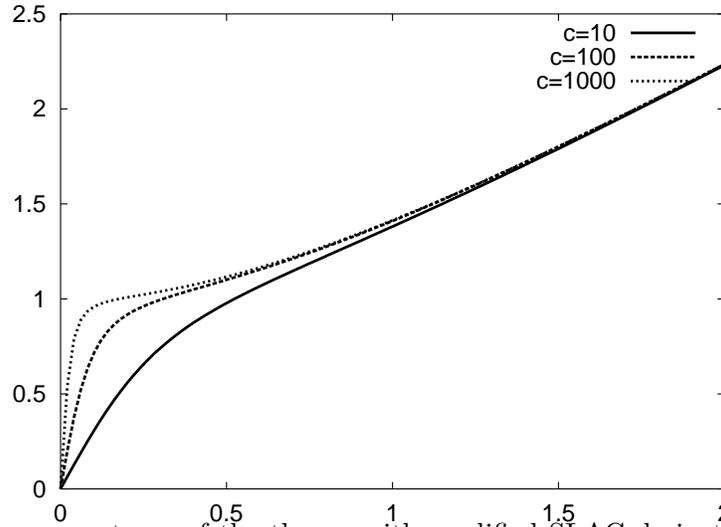}
    \hfill
   \parbox{14.cm}{
    \caption[]{The energy spectrum of the theory with modified SLAC
derivative for several values of $c$. }}
  \end{center}
\label{goldstone}
\end{figure*}

\begin{figure*}
\begin{center}
    \leavevmode
    \epsfxsize=10.cm
    \epsffile[58 57 401 293 ]{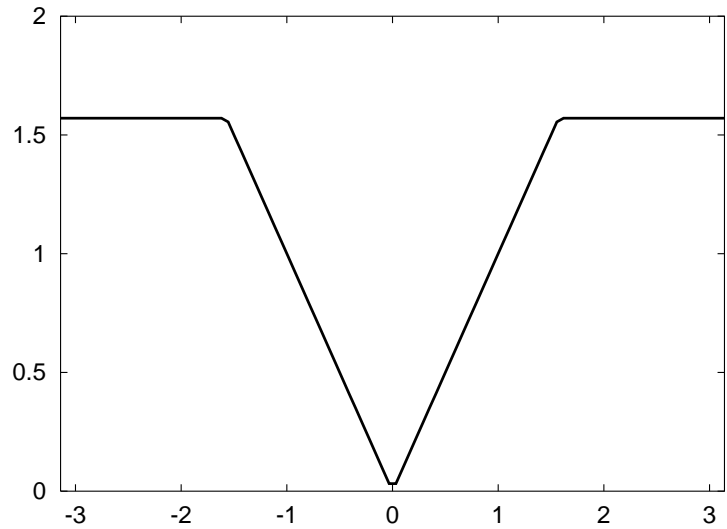}
    \hfill
   \parbox{14.cm}{
    \caption[]{The energies of one-particle states 
 (in units of $1/a$)
for
 modified Wilson derivative.
}}
  \end{center}
\label{wilsonmod}
\end{figure*}




\end{document}